\documentclass[conference]{IEEEtran}
\usepackage{cite}
\usepackage{amsmath,amssymb,amsfonts}
\usepackage{algorithmic}
\usepackage{graphicx}
\usepackage{textcomp}
\usepackage{xcolor}
\def\BibTeX{{\rm B\kern-.05em{\sc i\kern-.025em b}\kern-.08em
    T\kern-.1667em\lower.7ex\hbox{E}\kern-.125emX}}
\makeatletter
\newcommand{\newlineauthors}{%
  \end{@IEEEauthorhalign}\hfill\mbox{}\par
  \mbox{}\hfill\begin{@IEEEauthorhalign}
}
\makeatother
\newcommand{\imagepath}{./}

\newcommand{\code}{\mathbf{x}_t}
\newcommand{\prev}{\mathbf{x}_{t-1}}
\newcommand{\entropy}{\mathrm{H}}
\newcommand{\refframe}{\hat{\mathbf{x}}_{t-1}}
\newcommand{\balph}{\boldsymbol{\alpha}}
\newcommand{\sysoutput}{\hat{\mathbf{x}}_t}
\newcommand{\prediction}{\tilde{\mathbf{x}}_t}
\newcommand{\diagramscale}{0.39}
\newcommand{\copyarea}{\mathcal{S}}
\newcommand{\noncopyarea}{\bar{\copyarea}}
\newcommand{\flow}{\mathbf{v}}
\newcommand{\warping}{w}
\usepackage{float}
\usepackage{subcaption}
\usepackage{multirow}
\usepackage{pgfplots, soul} % PGFPlots charge automatiquement TikZ
\begin{document}

% Copyright notices.
% ------------------
% Select one of the four copyright notices below. Only required for the camera-ready paper submission.

% * For all other papers:
\IEEEoverridecommandlockouts
\IEEEpubid{\makebox[\columnwidth]{978-1-7281-9320-5/20/\$31.00 {\copyright}2020 IEEE \hfill} \hspace{\columnsep}\makebox[\columnwidth]{ }}

\title{Optical Flow and Mode Selection for Learning-based Video Coding}

\author{\IEEEauthorblockN{Th\'{e}o Ladune}
\IEEEauthorblockA{\textit{Orange}\\
Rennes, France\\
\texttt{\footnotesize theo.ladune@orange.com}}
\and
\IEEEauthorblockN{Pierrick Philippe}
\IEEEauthorblockA{\textit{Orange}\\
Rennes, France\\
\texttt{\footnotesize pierrick.philippe@orange.com}}
\newlineauthors
\IEEEauthorblockN{Wassim Hamidouche}
\IEEEauthorblockA{\textit{Univ. Rennes, INSA Rennes} \\
\textit{CNRS, IETR -- UMR 6164}\\
Rennes, France \\
\texttt{\footnotesize wassim.hamidouche@insa-rennes.fr}}
\and
\IEEEauthorblockN{Lu Zhang}
\IEEEauthorblockA{\textit{Univ. Rennes, INSA Rennes} \\
\textit{CNRS, IETR -- UMR 6164}\\
Rennes, France \\
\texttt{\footnotesize lu.ge@insa-rennes.fr}}
\and
\IEEEauthorblockN{Olivier D\'{e}forges}
\IEEEauthorblockA{\textit{Univ. Rennes, INSA Rennes} \\
\textit{CNRS, IETR -- UMR 6164}\\
Rennes, France \\
\texttt{\footnotesize olivier.deforges@insa-rennes.fr}}
}

\maketitle
\IEEEpubidadjcol
\begin{abstract}
This paper introduces a new method for inter-frame coding based on two
complementary autoencoders: MOFNet and CodecNet. MOFNet aims at computing
and conveying the Optical Flow and a pixel-wise coding Mode selection. The
optical flow is used to perform a prediction of the frame to code. The
coding mode selection enables competition between direct copy of the prediction or
transmission through CodecNet.

The proposed coding scheme is assessed under the \textit{Challenge on Learned
Image Compression 2020} (CLIC20) P-frame coding conditions, where it
is shown to perform on par with the state-of-the-art video codec ITU/MPEG HEVC.
Moreover, the possibility of copying the prediction enables to learn the optical
flow in an end-to-end fashion \textit{i.e.} without relying on
pre-training and/or a dedicated loss term.
\end{abstract}


\section{Introduction and Related Works}

% Video signal
Video signals exhibit a high level of redundancies, leveraged by compression
systems to reduce the transmission rate. Those redundancies can be classified
into two categories, spatial or temporal. Classical video compression systems
such as ITU/MPEG (AVC \cite{DBLP:journals/cm/MarpeWS06}, HEVC
\cite{Sullivan:2012:OHE:2709080.2709221} and VVC \cite{VVC_Ref}) codecs reduce
temporal redundancies through motion compensation. It relies on motion vectors,
representing motion between reference frames (available at the decoder) and the
current frame, which are estimated and conveyed as side-information. Motion
vectors are used to perform a prediction of the current frame, allowing the
system to transmit only the prediction error \textit{i.e.} difference between
the signal and its prediction (the residue), lowering the required rate. A frame
coded without temporal dependency is called an \textit{intra} frame in contrast
to an \textit{inter} frame relying on information from other frames.

% Motion compensation
Inspired by traditional codecs, most neural network-based video coding
approaches
\cite{DBLP:conf/cvpr/LuO0ZCG19,DBLP:journals/corr/abs-1912-06348,Djelouah_2019_ICCV,DBLP:journals/corr/abs-2003-01966}
also rely on motion compensation for inter frame processing. These methods use
an optical flow network (such as SpyFlow \cite{DBLP:journals/corr/RanjanB16} or
PWC-Net \cite{DBLP:conf/cvpr/SunY0K18}) to compute pixel-wise motion vectors.
Motion vectors are transmitted by a dedicated neural-based coding system and
used for motion compensation. The prediction is exploited through a simple
encoding of the prediction error (difference between the frame and its
prediction), computed either in image
\cite{DBLP:conf/cvpr/LuO0ZCG19,DBLP:journals/corr/abs-1912-06348} or in latent
domain \cite{Djelouah_2019_ICCV}. As stated in
\cite{DBLP:journals/corr/abs-2004-04342}, it is not trivial to learn the optical
flow with a loss function only based on the RD-cost. Consequently, previous
work relies either on pre-trained network or on a dedicated loss term during
training, resulting in a cumbersome training process.

% Our work 
% Simple, off-the-shelf block
In this work a method for inter frame coding is introduced. This method is based
on two autoencoder neural networks. First, a mode selection and optical flow
estimation network (MOFNet) is proposed. The role of MOFNet is to compute and convey the
optical flow and additionally a pixel-wise coding mode selection. MOFNet
arbitrates each pixel between copy from the prediction (\textit{Skip Mode} in
classical codecs) or transmission through the coding network CodecNet. Inspired
by the approach proposed in \cite{DBLP:journals/corr/abs-2004-04342}, CodecNet
learns the appropriate mixture of the current frame and its prediction, allowing
to exploit more information than direct residual coding.

MOFNet is the key component of the proposed method. Similarly to traditional
codecs, it permits competition between coding modes, improving the whole coding
scheme performances by compensating CodecNet potential weaknesses. The
availability of skip mode enables to learn the optical flow in an end-to-end
fashion, without relying on separate training or a dedicated loss term,
overcoming an issue of existing methods.

The proposed method benefits are illustrated under the \textit{Challenge on
Learned Image Compression 2020} (CLIC20) P-frame coding the test conditions
\cite{CLIC20_web_page}. It is shown to achieve state-of-the-art performance,
performing on par with HEVC.

\section{Problem Formulation}

This section introduces the general task of P-frame coding and narrows it down
to the CLIC20 test conditions.

Let $\mathcal{V} = \left\{\mathbf{x}_i\right\}_{i \in \mathbb{N}}$ be a video,
represented as a set of frames, with each frame $\mathbf{x}_i \in \mathbb{R}^{C
\times H \times W}$ where $C$, $H$ and $W$ denote the number of color channels,
height and width of the frame, respectively. This work targets a P-frame
coding, which consists in coding the current frame $\code$ with previous frames
${\mathbf{x}_{< t} = \left\{\mathbf{x}_{t-1}, \mathbf{x}_{t-2},\ldots \right\}}$
already transmitted and available at decoder side to be used as references
${\hat{\mathbf{x}}_{< t} = \left\{\hat{\mathbf{x}}_{t-1},
\hat{\mathbf{x}}_{t-2},\ldots \right\}}$. In order to reduce temporal
redundancies, a prediction $\prediction$ of $\code$ is made available, based on
${\hat{\mathbf{x}}_{< t}}$ and side-information (such as motion). 

In this work a lossy P-frame coding scheme is considered through a rate-distortion (RD) trade-off:
\begin{equation}
    \mathcal{L}(\lambda) = \mathrm{D}(\sysoutput, \code) + \lambda\mathrm{R},\ \text{with}\ \sysoutput = s(\prediction, \code),
\end{equation}
where $\mathrm{D}$ is a distortion measure, $\sysoutput$ is the reconstruction
from a coding scheme $s$ with an associated rate $\mathrm{R}$ weighted by a Lagrange
multiplier $\lambda$. Following the CLIC 20 P-frame coding test conditions, the
distortion measure is based on the Multi Scale Structural Similarity Metric
(MS-SSIM)\cite{Wang03multi-scalestructural}:
\begin{equation*}
\mathrm{D}(\sysoutput, \code) = 1 - \text{MS-SSIM}(\sysoutput, \code).
\end{equation*}

The CLIC20 P-frame challenge asusumes that there is only one reference frame
available, \textit{i.e.} ${\hat{\mathbf{x}}_{< t} = \refframe}$, whose coding is
supposed to be lossless ($\refframe = \prev$).

\section{Proposed method}
\label{sec:proposed_method}
\begin{figure}
    \centering
    \includegraphics[width=\columnwidth]{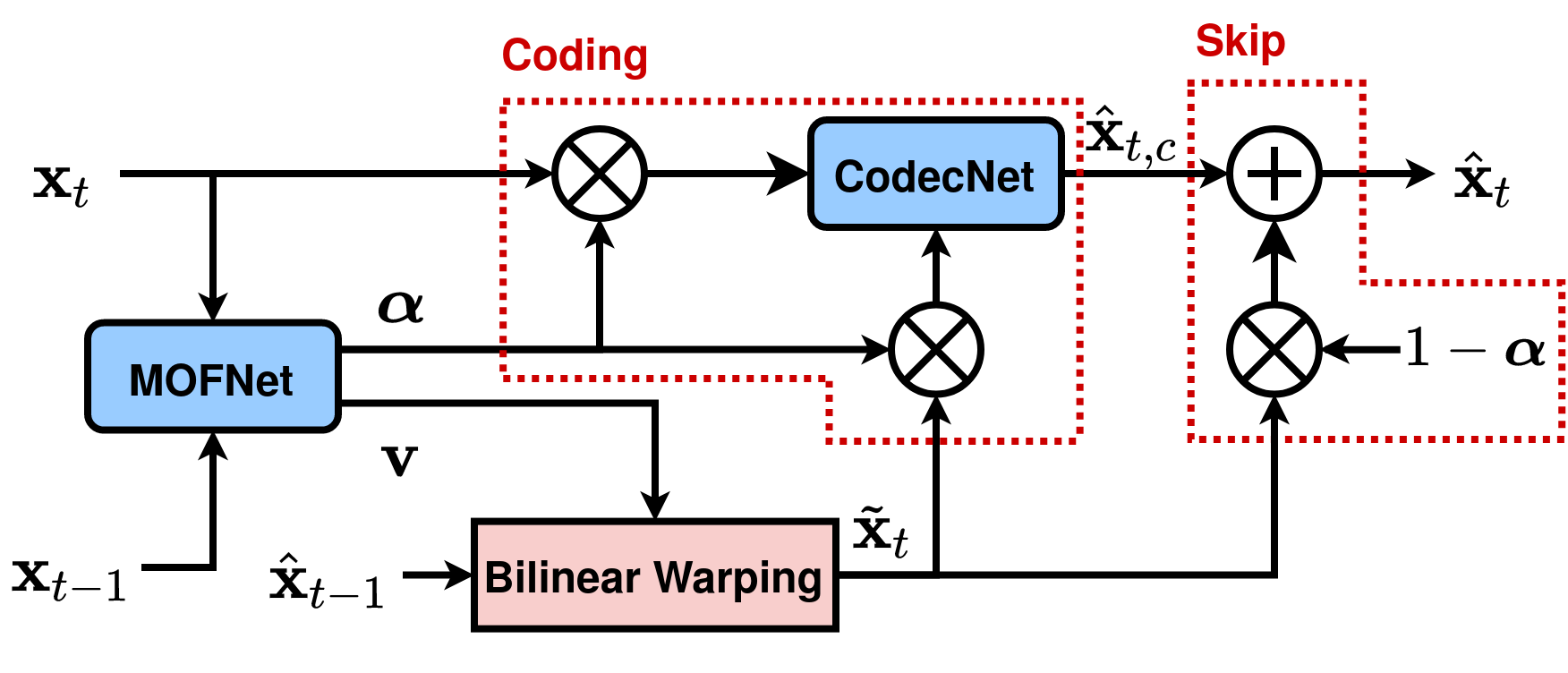}
    \caption{Architecture of the proposed system.}
    \label{CompleteSystemDiagrams}
\end{figure}

This section details the main components of the proposed coding scheme,
presented in Fig. \ref{CompleteSystemDiagrams}.

% ===== ARCHITECTURE FIGURES ===== %
\begin{figure*}[htb]
    %\centering % Not needed
    \begin{subfigure}[b]{1\columnwidth}
        \centering
        \includegraphics[scale=0.17]{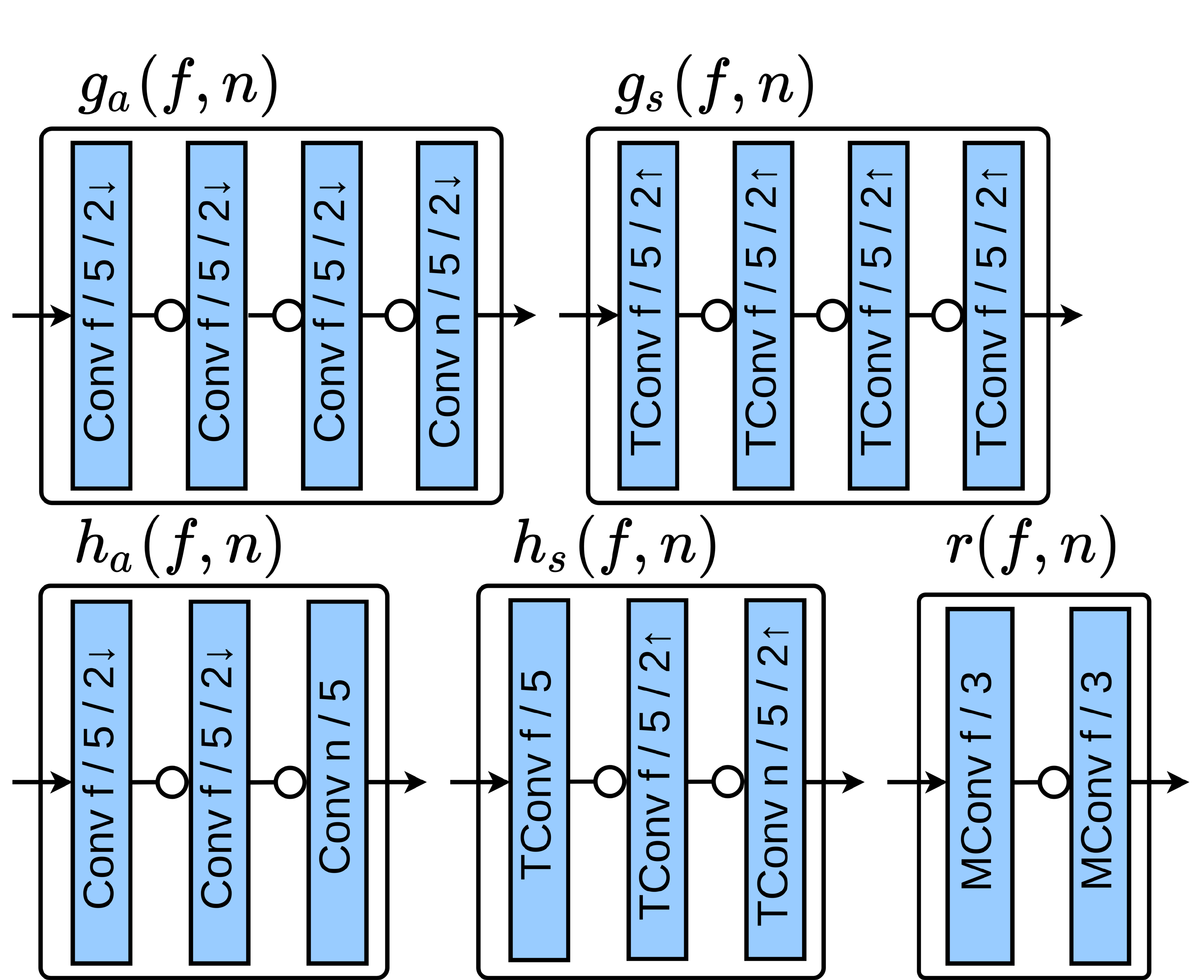}
        \caption{Basic building blocks of the proposed systems. $f$ and $n$
        respectively stand for the number of internal and output features.
        Rounded arrows denote non-linearities. Convolutions parameters are filters
        number $\times$ kernel size / stride. TConv and MConv stand respectively
        for Transposed convolution and Masked convolution.}
        \label{RawBlocksDiagrams}
    \end{subfigure}
    \hfill
    \begin{subfigure}[b]{1\columnwidth}
        \centering
        \raisebox{1.5cm}{\includegraphics[scale=\diagramscale]{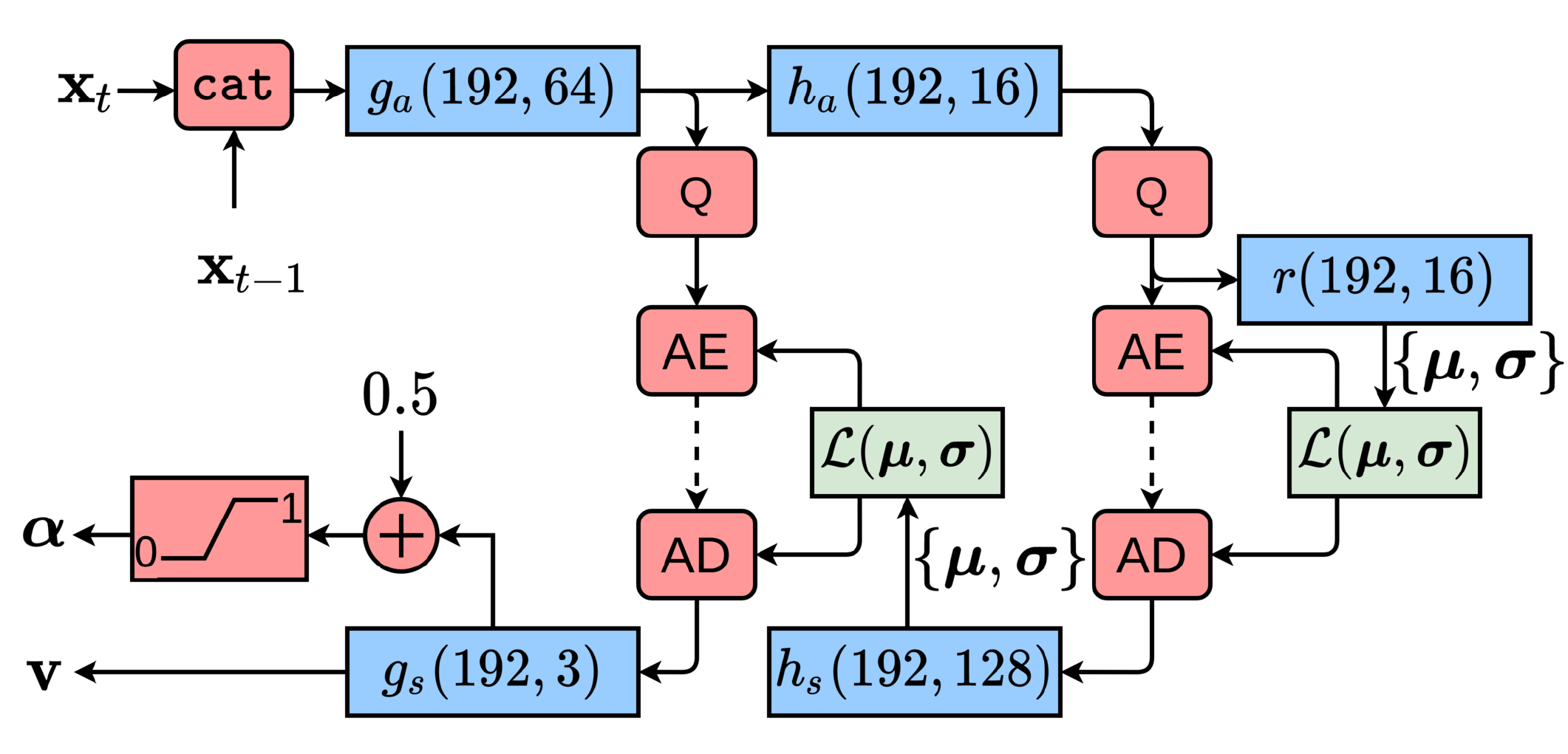}}
        \caption{MOFNet architecture. All components use LeakyReLU.}
        \label{MOFNetDiagram}
    \end{subfigure}
    %% leave a blank line to create a line break
    \begin{subfigure}[b]{1\columnwidth}
        \vspace{0.4cm}
        \centering
        \includegraphics[scale=\diagramscale]{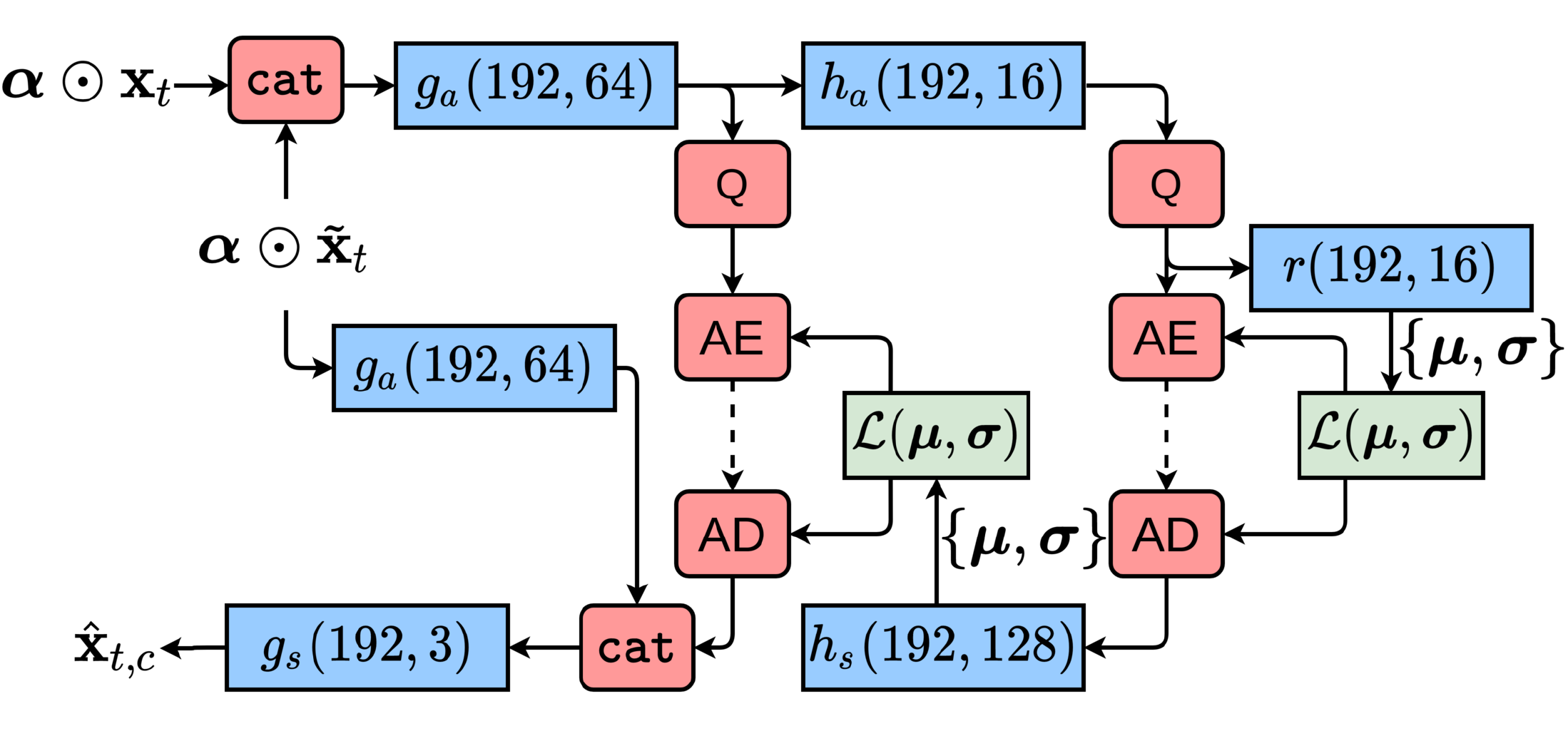}
        \caption{CodecNet architecture. $g_a$ and $g_s$ use GDN
        \cite{DBLP:conf/iclr/BalleLS17}, $h_a$, $h_s$ and $r$ use LeakyReLU.}
        \label{CodecNetConditionalDiagram}
    \end{subfigure}
    \hfill
    \begin{subfigure}[b]{1\columnwidth}
        \centering
        \raisebox{0.2cm}{\includegraphics[scale=\diagramscale]{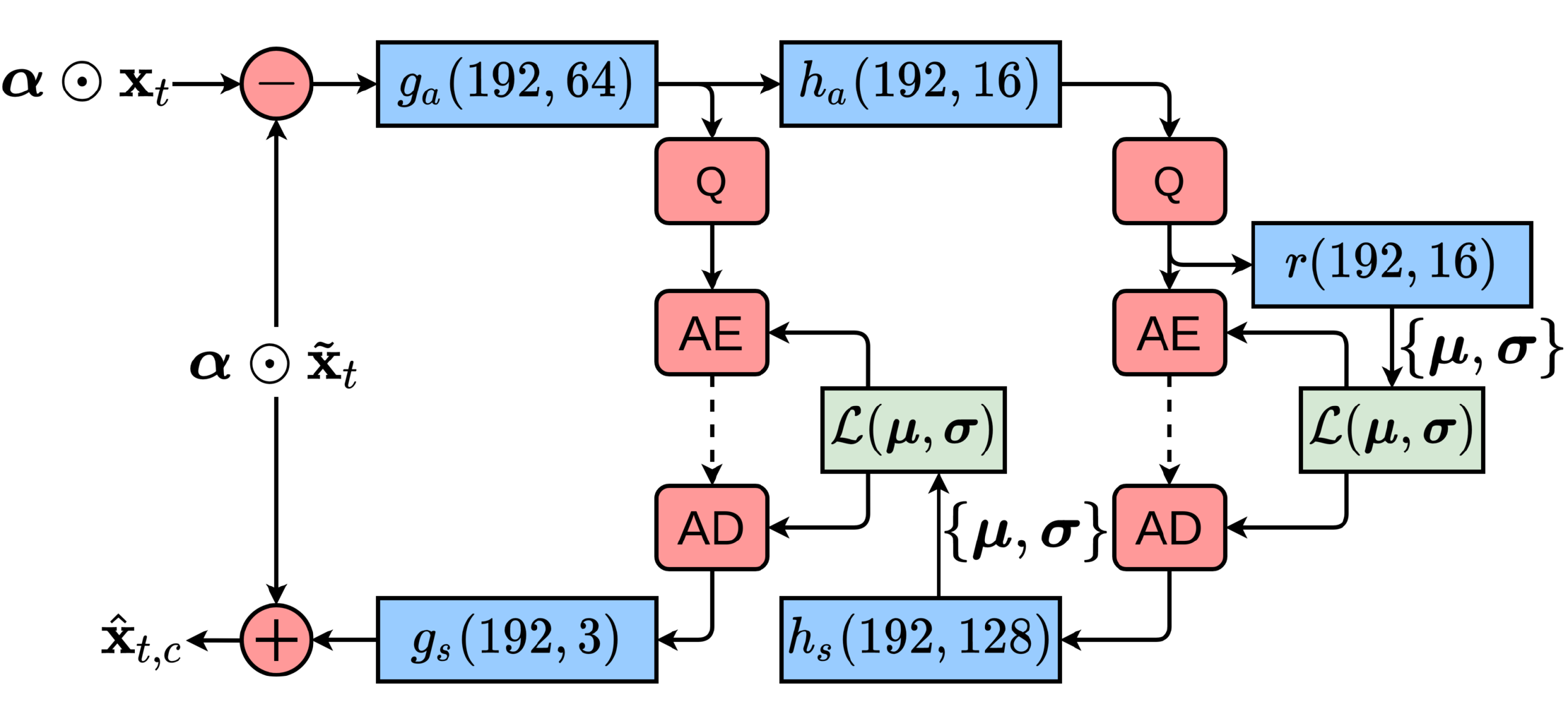}}
        \caption{Residual coding architecture used for ablation study, section
        \ref{sec:ablation}. $g_a$ and $g_s$ use GDN
        \cite{DBLP:conf/iclr/BalleLS17}, $h_a$, $h_s$ and $r$ use LeakyReLU.}
        \label{CodecNetDiffDiagram}
    \end{subfigure}
    \caption{Detailed architecture of all proposed networks. $g_a$ and $g_s$ are
    the main encoder/decoder, $h_a$ and $h_s$ are the hyperprior encoder/decoder
    and $r$ is an auto-regressive module as in \cite{DBLP:conf/nips/MinnenBT18}.
    There is no weight sharing among transforms denoted by the same function.
    $\texttt{cat}$ stands for concatenation along the feature axis, Q for
    quantization, AE and AD for arithmetic encoding/decoding with a Laplace
    distribution $\mathcal{L}$.}
    \label{fig:architecture}
\end{figure*}
% ===== ARCHITECTURE FIGURES ===== %

\subsection{MOFNet: Mode Selection and Optical Flow Estimation}

Performing a proper prediction of the current frame is an essential element of
video coding systems. Indeed, most parts of the frame $\code$ can be
recovered from already received frames ${\hat{\mathbf{x}}_{< t}}$ using motion
vectors transmitted at low rate. 

In this work, a dense optical flow ${\flow \in \mathbb{R}^{2 \times H \times
W}}$ is used to represent the 2-D motion of each pixel between $\refframe$ and $\code$.
The estimated optical flow is used to perform the prediction:

\begin{equation}
    \prediction = \warping(\refframe, \flow),
\end{equation}

where $\warping$ is a bilinear warping, as illustrated in Fig. \ref{CompleteSystemDiagrams}.

The proposed coding scheme splits $\code$ into two complementary pixels sets
$\copyarea$ and $\noncopyarea$, corresponding to two coding modes. The pixels in
$\copyarea$ are directly copied from the prediction $\prediction$ as
\textit{Skip Mode} in classical codecs. Those in $\noncopyarea$ are transmitted
by an autoencoder. The presence of two competiting coding modes allows to select
the most suited one for each pixel, resulting in better RD performances.
However, this partitioning into two sets is not straightforward, as the rate and
the distortion of a pixel depends on the coding choice made for both previous
and future pixels. 

A single network MOFNet is proposed, to compute and convey the coding mode
selection and the flow estimation. MOFNet is defined as a function $m$:
\begin{equation}
    \mathrm{R}_{m},\ \balph,\ \flow = m\left(\prev,\code\right),
\end{equation}
where ${\balph \in \left[0, 1\right]^{H \times W}}$ is the pixel-wise
weighting matrix, ${\flow}$ the optical flow and $\mathrm{R}_m$ the associated rate. The
pixel-wise weighting matrix $\balph$ is real-valued such that smooth transitions
between coding modes are possible, avoiding blocking artifacts.

\subsection{CodecNet}

An immediate way of using the prediction is to perform residual coding
\textit{i.e.} coding only the prediction error $\code - \prediction$. Albeit
widely used in legacy video coding systems, this method is not the
best option for leveraging information from $\prediction$. Indeed, from a source
coding perspective: 
\begin{equation}
    \entropy(\code \mid \prediction) \leq  \entropy(\code - \prediction),
    \label{eq:entropy}
\end{equation}
where $\entropy$ denotes the Shannon entropy. Therefore coding $\code$ while
retrieving all information from $\prediction$ can result in less information to
transmit than residual coding.

In this work, an autoencoder CodecNet is used to transmit $\noncopyarea$,
selected by the pixel-wise weighting matrix $\balph$. CodecNet learns the appropriate
mixture of $\code$ and $\prediction$ for both the encoder and the decoder,
resulting in potentially better coding performances than direct residual coding.
In contrast with residual coding, the processing performed by CodecNet is
denoted as \textit{conditional coding} in the remaining of the paper. CodecNet
is defined as a function $c$, coding $\code$ using information from
$\prediction$:
\begin{equation}
    \mathrm{R}_{c},\ \hat{\mathbf{x}}_{t, c} = c\left(\alpha \odot \prediction, \alpha \odot \code\right),
\end{equation}
where element-wise matrix multiplication is denoted by $\odot$,
$\hat{\mathbf{x}}_{t, c} \in \mathbb{R}^{C \times H \times W}$ is the
reconstruction of $\alpha \odot \code$ and $\mathrm{R}_c$ the associated rate.
The same $\balph$ is used for all $C$ color channels. 

\subsection{Complete System}

One of MOFNet purposes is to split $\code$ transmission between CodecNet and
skip mode. Thus the complete reconstruction is:

\begin{equation}
    \sysoutput = \underbrace{(1 - \balph) \odot \prediction}_{\text{Skip}} +
    \underbrace{c(\balph \odot \prediction, \balph \odot \code)}_{\text{Conditional coding}}.
    \label{eq:sysoutput}
\end{equation}
This equation highlights that the role of $\balph$ is to zero areas from $\code$
before coding them with CodecNet, in order to save their associated rate.
Figures \ref{ex:ModeNetOutputAlpha}, \ref{ex:CodecNetPart} and
\ref{ex:CodecNetRate} illustrate that CodecNet does not allocate bits to areas
zeroed by $\balph$. MOFNet and CodecNet are trained in an end-to-end fashion to
minimize the rate-distortion trade-off:

\begin{equation}
    \mathcal{L}(\lambda) = \mathrm{D}\left(\sysoutput, \code\right) + \lambda \left(\mathrm{R}_m + \mathrm{R}_c\right).
    \label{eq:globalLoss}
\end{equation}

% ===== EXAMPLE FIGURES ===== %
\newcommand{\examplewidth}{0.765\columnwidth}
\newcommand{\expath}{\imagepath}
\begin{figure*}[htb]
    % \centering % Not needed
    \begin{subfigure}[b]{1\columnwidth}
        \centering
        \includegraphics[width=\examplewidth]{\expath/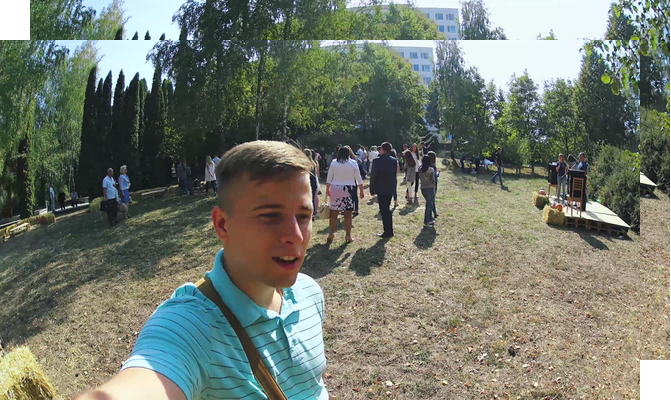}
        \caption{The pair of frames $(\prev,\code)$.}
        \label{ex:InputPairs}
    \end{subfigure}
    \hfill
    \begin{subfigure}[b]{1\columnwidth}
        \centering
        \includegraphics[width=\examplewidth]{\expath/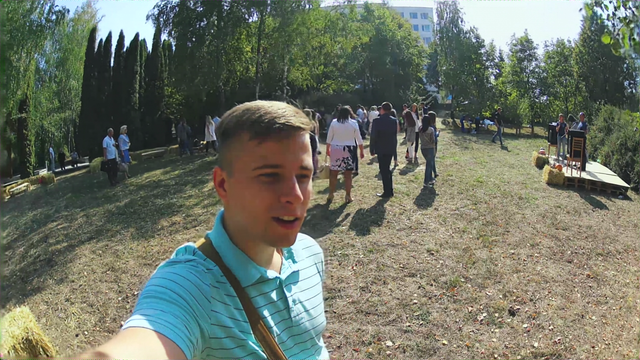}
        \caption{Reconstructed frame: $\sysoutput = (1 - \balph) \odot \prediction + c(\balph \odot \prediction, \balph \odot \code)$.}
        \label{ex:SystemOutput}
    \end{subfigure}      
    %% leave a blank line to create a line break
    
    \begin{subfigure}[b]{1\columnwidth}
        \centering
        \includegraphics[width=\examplewidth]{\expath/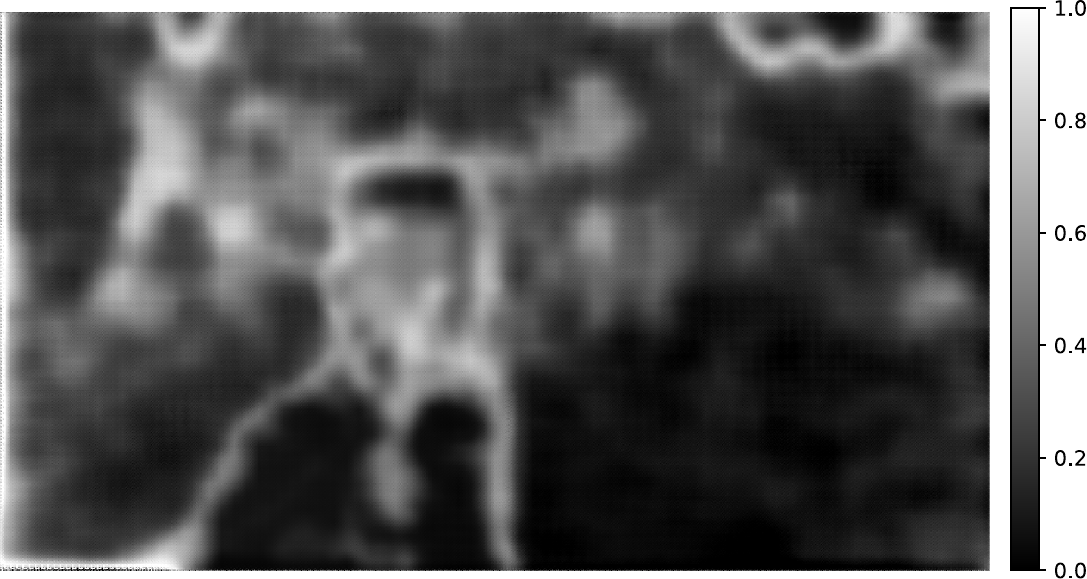}
        \caption{Coding mode selection matrix $\balph$. Black areas correspond to skip mode, white ones to CodecNet.}
        \label{ex:ModeNetOutputAlpha}
    \end{subfigure}        
    \hfill    
    \begin{subfigure}[b]{1\columnwidth}
        \centering
        \raisebox{0.5cm}{\includegraphics[width=\examplewidth]{\expath/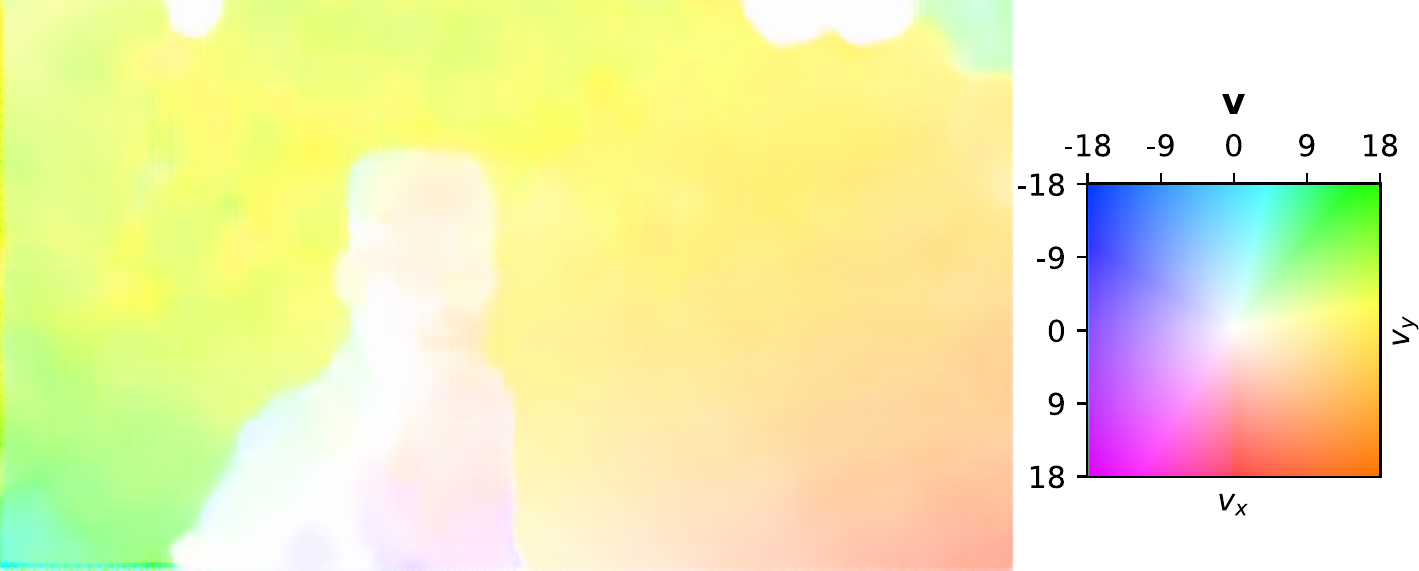}}
        \caption{Optical flow $\flow$. Displacements are in pixels.\\ \hphantom{a}}
        \label{ex:flow}
    \end{subfigure}
    %% leave a blank line to create a line break

    \begin{subfigure}[b]{1\columnwidth}
        \centering
        \includegraphics[width=\examplewidth]{\expath/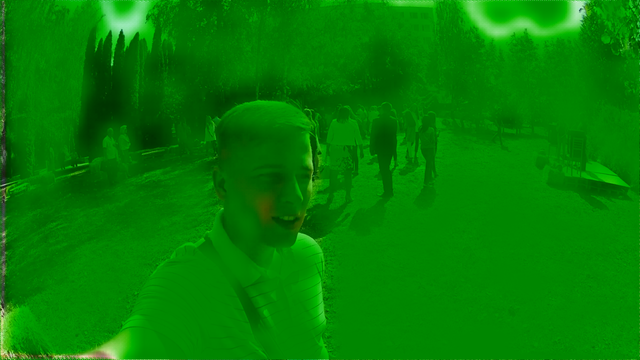}
        \caption{Areas selected for the CodecNet: $\balph \odot \code$.}
        \label{ex:CodecNetPart}
    \end{subfigure}
    \hfill
    \begin{subfigure}[b]{1\columnwidth}
        \centering
        \includegraphics[width=\examplewidth]{\expath/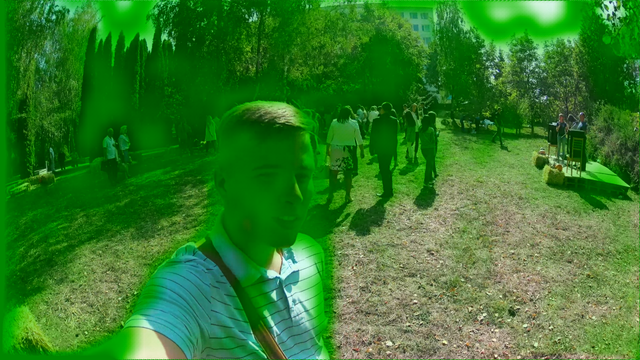}
        \caption{Areas selected for skip mode: $(1 - \balph) \odot \prediction$.}
        \label{ex:CopyPart}
    \end{subfigure}
    %% leave a blank line to create a line break

    \begin{subfigure}[b]{1\columnwidth}
        \centering
        \includegraphics[width=\examplewidth]{\expath/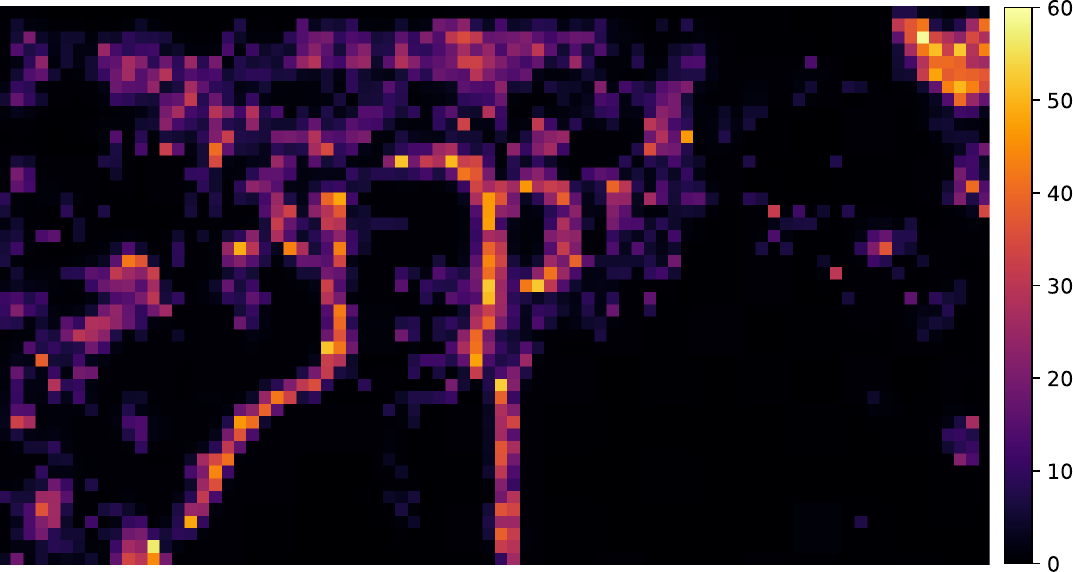}
        \caption{Spatial distribution of CodecNet rate in bits.}
        \label{ex:CodecNetRate}
    \end{subfigure}
    \hfill
    \begin{subfigure}[b]{1\columnwidth}
        \centering
        \includegraphics[width=\examplewidth]{\expath/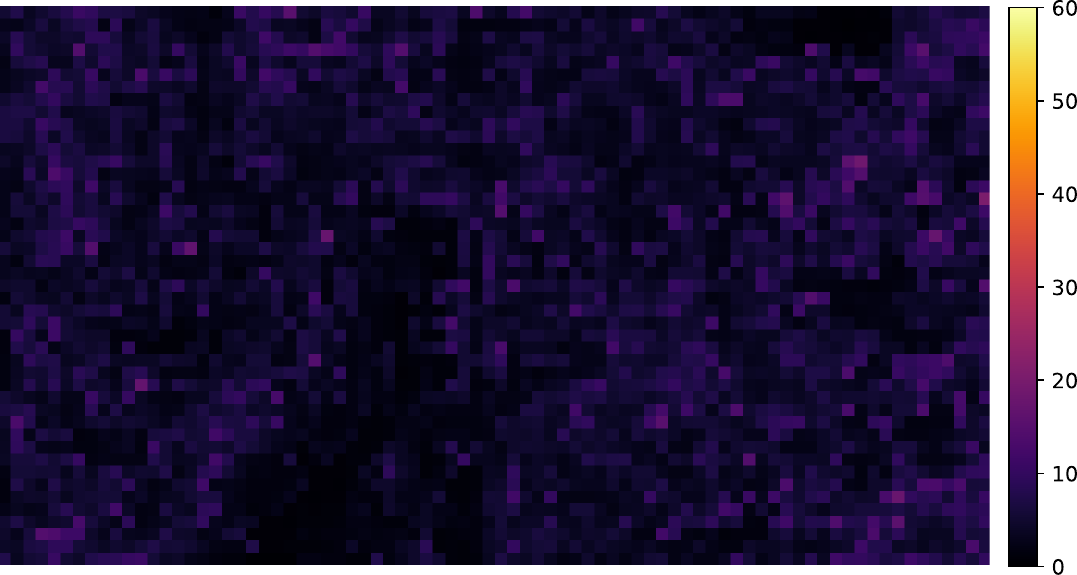}
        \caption{Spatial distribution of MOFNet rate in bits.}
        \label{ex:MOFNetRate}
    \end{subfigure}    
    \caption{Details of the system behavior. The pair of frames $(\prev,\code)$
    represents a static man with a rotating background. For this example,
    $\text{MS-SSIM} = 0.982$, $\mathrm{R}_c = 0.022$~bpp and $\mathrm{R}_m =
    0.019$~bpp.}
    \label{fig:Example}
\end{figure*}
% ===== EXAMPLE FIGURES ===== %

\section{Practical Implementation}

\subsection{Networks Architecture}

The two neural networks proposed in section \ref{sec:proposed_method}, MOFNet
and CodecNet, are described in Fig. \ref{fig:architecture}. They are both based
on the common autoencoder with hyperprior (AE-HP) architecture
\cite{DBLP:conf/iclr/BalleMSHJ18} used in previous learned image coding systems.

MOFNet role is to compute and convey the optical flow $\flow$ and the pixel-wise
weighting $\balph$. Authors in \cite{DBLP:journals/corr/abs-1912-06348} show
that a single network can perform both estimation and coding of $\flow$. This
work follows this method and uses a common learned image coding architecture,
depicted in Fig. \ref{MOFNetDiagram}. MOFNet takes $\prev$ and $\code$ as inputs
and retrieves $\flow$ and $\balph$ at the decoder side. To ensure that $\balph$ remains
in $\left[0, 1\right]$, a clipping function is used. A bias of $0.5$ is
added before clipping as it empirically ensures better convergence. 

The purpose of CodecNet is to transmit pixels $\noncopyarea$ of $\code$
conditioned to its prediction $\prediction$. It is designed as an AE-HP system
with the ability to learn an arbitrary complex mixture of $\code$ and
$\prediction$, at the encoder side and the decoder side. CodecNet architecture
(see Fig. \ref{CodecNetConditionalDiagram}) is a direct extension of image
coding autoencoders with both the frame and its prediction as inputs. Therefore,
the encoder is able to learn a non-linear mixture of $\code$ and $\prediction$.
The same principle is used for the decoder, which has the latents from $\code$
and $\prediction$ as input, allowing it to invert the transform performed by the
encoder.

\subsection{Training}

All networks are trained in an end-to-end fashion to minimize the global loss
function stated in eq. \eqref{eq:globalLoss}. Non-differentiable parts are
approximated as in Ball\'e's work
\cite{DBLP:conf/iclr/BalleLS17,DBLP:conf/iclr/BalleMSHJ18} to make the training
possible. 

To the best of our knowledge, all previous work learn the flow $\flow$ with
either a pre-trained network and/or a dedicated loss term. In the proposed
coding scheme, the optical flow can be learned without a seperately pre-trained
network or a dedicated additional loss term. Indeed, the areas directly copied
from $\prediction$ heavily foster the learning of a proper flow, with no need of
pre-training and/or a dedicated loss term.

However, due to the competition between signal paths, some care is taken when training.
The training process is composed of three phases:

\begin{enumerate}
    \item During the first five epochs, skip mode and CodecNet are
    not ready to compete. Thus, $\balph$ is frozen and set to 1 for one half of
    the frame, 0 for the other half. This allows to learn a meaningful MOFNet and
    CodecNet without interference between them.
    \item Alternate training of MOFNet and CodecNet, one
    epoch for each (\textit{i.e.} the other network weights are frozen) for 45 epochs.
    \item Joint training of MOFNet and CodecNet for 20 epochs.
\end{enumerate}

Training is performed on the CLIC20 P-frame dataset \cite{CLIC20_web_page}. The
training set is composed of half a million $256 \times 256$ pairs of crops,
randomly extracted from consecutive frames. The same learning rate of $10^{-4}$
is used for all three phases with a decrease down to $4 \times 10^{-6}$ during the
final phase.

\section{System Behavior and Visualisation}

The processing of a pair of frames  $(\prev,\code)$ is thoroughly described in
this section, illustrated in Fig. \ref{fig:Example}. The example frames are
extracted from the CLIC20 P-frame dataset, sequence \textit{Vlog\_2160P-310b}
frames 36 and 37.

First, MOFNet takes $\prev$ and $\code$ (shown in Fig. \ref{ex:InputPairs}) as
inputs. The pair of frames are encoded and decoded as $\flow$ and $\balph$. The
optical flow $\flow$ (illustrated in Fig. \ref{ex:flow}) is used to perform a
prediction $\prediction$ of $\code$ through a bilinear warping. Then, the
pixel-wise weighting $\balph$  (see Fig. \ref{ex:ModeNetOutputAlpha}) arbitrates
between skip mode and CodecNet. Fig. \ref{ex:CodecNetPart} and \ref{ex:CopyPart}
present the areas selected for both coding modes\footnote{As images are in YCbCr
format, zeroed areas appear in green}. Finally, the two coding modes are
combined to obtain the reconstructed frame, shown in Fig. \ref{ex:SystemOutput}.

$\copyarea$ represent areas in $(\prev, \code)$ more suited for skip mode than
coding, \textit{i.e.} areas which are either well handled by motion compensation
or too costly to transmit. In order to select these areas for skip mode,
$\balph$ tends to be zero for pixels in $\copyarea$. These areas correspond to
the green ones in Fig. \ref{ex:CodecNetPart}, \textit{e.g.} the grass and most of
the man. By contrast, $\balph$ is close to one for pixels in $\noncopyarea$,
which are not well predicted enough or relatively easy to transmit. To achieve
an acceptable quality, those pixels rely on transmission by CodecNet. These
areas appear in green in Fig. \ref{ex:CopyPart}. They correspond to contents
which are difficult to predict such as the edges of the man or the leaves of the
tree. 

Figures \ref{ex:CodecNetRate} and \ref{ex:MOFNetRate} represent the spatial
distribution of the rate of CodecNet and ModeNet. As expected from eq.
\eqref{eq:sysoutput}, areas with a small $\balph$ are zeroed before CodecNet
and thus transmitted for free. The motion and the partitioning conveyed by
MOFNet is complex throughout the frame, resulting in a small rate
almost evenly distributed spatially.

This illustration highlights that MOFNet is able to learn a complex optical
field, \textit{e.g.} modeling a rotation motion for the background while not
including the man in the foreground. In the meantime, MOFNet is also able to
learn $\balph$, an accurate and smooth partitioning of the frame, which
indicates the properly predicted areas and those needing to rely on CodecNet.
Both $\flow$ and $\balph$ are conveyed at low-bitrate (around 0.02~bpp in this
example).

\section{Experimental Results}

% ======= USED FOR RESULTS PLOT ======= %
\definecolor{graphblue}{rgb}{0.2, 0.2, 0.6}
\definecolor{graphyellow}{rgb}{0.953, 0.654, 0.071}
\definecolor{graphred}{rgb}{0.8, 0.0, 0.0}
\definecolor{graphgreen}{rgb}{0.07, 0.53, 0.03}
\definecolor{graphpurple}{rgb}{0.41, 0.21, 0.61}
\newcommand{\markscale}{1}
\newcommand{\markstarscale}{1.3}
\newcommand{\hevcmark}{star}
\newcommand{\fullsystemmark}{square*}
\newcommand{\ablationmarklinestyle}{dashed}
\newcommand{\ablationmark}{o}

\newcommand{\resgraphwidth}{9cm}
\newcommand{\resgraphheigth}{7.8cm}
\newcommand{\resgraphxmin}{0}
\newcommand{\resgraphxmax}{0.2}
\newcommand{\resgraphymin}{16}
\newcommand{\resgraphymax}{26}

\definecolor{battleshipgrey}{rgb}{0.52, 0.52, 0.51}
\definecolor{davysgrey}{rgb}{0.33, 0.33, 0.33}
\definecolor{ashgrey}{rgb}{0.7, 0.75, 0.71}
\pgfplotsset{minor grid style={solid,battleshipgrey, dotted}}
\pgfplotsset{major grid style={solid, thick}}
% ======= USED FOR RESULTS PLOT ======= %

% ======== GLOBAL RESULTS PLOT ======== %
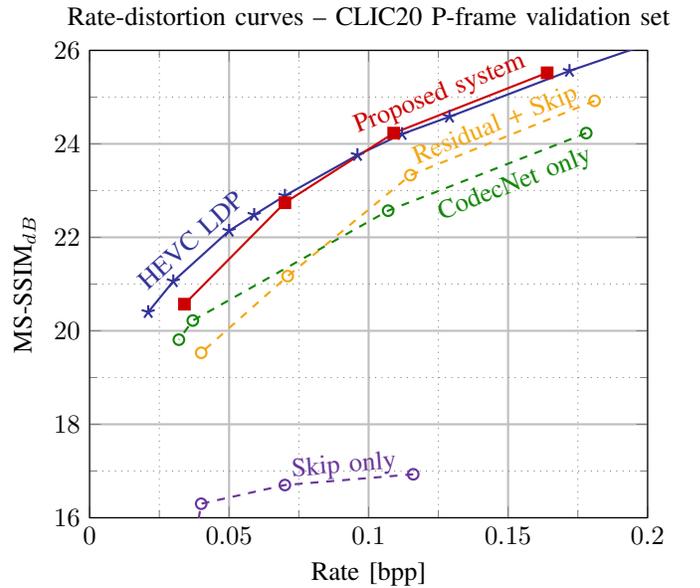
\begin{figure}[htb]
    \centering
    \begin{tikzpicture}
        \begin{axis}[
            grid= both ,
            xlabel = {Rate [bpp]} ,
            ylabel = {$\text{MS-SSIM}_{dB}$} ,
            xmin = \resgraphxmin, xmax = \resgraphxmax,
            ymin = \resgraphymin, ymax = \resgraphymax,
            ylabel near ticks,
            xlabel near ticks,
            width=\resgraphwidth,
            height=\resgraphheigth,
            x tick label style={
                /pgf/number format/.cd,
                fixed,
                % fixed zerofill,
                precision=2
                },
            xtick distance={0.05},
            ytick distance={2},
            minor y tick num=1,
            minor x tick num=1,
            title=Rate-distortion curves -- CLIC20 P-frame validation set,
        ]
            % HEVC LP
            \addplot[solid, thick, graphblue, mark=\hevcmark, mark options={solid, scale=\markstarscale}] coordinates{
                (0.021,20.40)
                (0.030,21.06)
                (0.050,22.14)
                (0.059,22.48)
                (0.070,22.89)
                (0.096,23.76)
                (0.112,24.21)
                (0.129,24.58)
                (0.172,25.56)
                (0.197,26.07)
                (0.226,26.62)
            }node [pos=0.2, sloped, anchor=south] {HEVC LDP};

            % Proposed System
            \addplot[thick, solid, graphred, mark=\fullsystemmark, mark options={solid, scale=\markscale}] coordinates {
                (0.164,25.52)
                (0.109,24.23)
                (0.070,22.74)
                % (0.043,21.08)
                (0.034,20.57)
            }node [pos=0.17, sloped, anchor=south, yshift=-0.03cm] {Proposed system};
        
            % No copy
            \addplot[\ablationmarklinestyle, thick, graphgreen, mark=\ablationmark, mark options={solid, scale=\markscale}] coordinates {
                (0.178,24.23)
                (0.107,22.57)
                % (0.073,20.85)
                (0.037,20.22)
                (0.032,19.81)
            }node [pos=0.15, sloped, anchor=north] {CodecNet only};
        
            % No coder
            \addplot[\ablationmarklinestyle, thick, graphpurple, mark=\ablationmark, mark options={solid, scale=\markscale}] coordinates {
                (0.116,16.93)
                (0.070,16.70)
                (0.040,16.30)
                (0.024,10.71)                
            }node [pos=0.02, sloped, anchor=south, yshift=-0.1cm] {Skip only};

            % Diff
            \addplot[\ablationmarklinestyle, thick, graphyellow, mark=\ablationmark, mark options={solid, scale=\markscale}] coordinates {
                (0.181,24.92)
                (0.115,23.33)
                (0.071,21.17)
                (0.040,19.53)
                % (0.034,16.93)
            }node [pos=0.15, sloped, anchor=south, yshift=-0.1cm] {Residual + Skip};        
        \end{axis}
    \end{tikzpicture}
    \caption{Rate-distortion performance of the systems, evaluated on CLIC20
    P-frame validation dataset. Quality metric is $\text{MS-SSIM}_{dB} = -10
    \log_{10} (1 - \text{MS-SSIM})$ (the higher the better). Rate is indicated
    in bits per pixel (bpp). Uncomplete systems used for ablation study are in dashed lines.}
    \label{fig:results}
\end{figure}
% ======== GLOBAL RESULTS PLOT ======== %

\subsection{System Performance}
\label{sec:perf}

The performance of the proposed inter frame coding scheme is assessed on the
CLIC20 validation set, under the challenge test conditions. In order to obtain a
RD-curve, the system is trained with different $\lambda$. The rate-distortion
curves are shown Fig.~\ref{fig:results}.

The proposed method is evaluated against the state-of-the-art video coder HEVC
in low-delay P (LDP) coding configuration. HEVC encodings are performed with the
HM 16.20 reference software slightly modified to be aligned with CLIC20 test
conditions where the reference frame is lossless. Our approach performs as good
as HEVC, proving the relevance of the proposed method. This demonstrates that
the optical flow learned in an end-to-end fashion, without pre-training or a
dedicated loss term, is able to achieve a temporal prediction competitive with
state-of-the-art motion compensation.

\subsection{Ablation study}
\label{sec:ablation}
The benefits of each component of the proposed system is also assessed in Fig.
\ref{fig:results}. In order to estimate the rate saving offered by the different
components, the BD-rate \cite{Bjontegaard} metric is used. It represents the
rate difference necessary to obtain identical quality between two systems.

The interest of performing a conditional coding of $\code$ and $\prediction$
instead of residual coding is evaluated by training a complete systems
(\textit{i.e.} including skip mode) while substituting CodecNet by a
neural-based residual codec, detailed in Fig. \ref{CodecNetDiffDiagram}. Its
performance is presented on Fig. \ref{fig:results} as \textit{Residual~+~Skip}.
According to the BD-rate metric, conditional coding reduces the rate by 32~\%
compared to direct residual coding, highlighting its relevance.

The improvements brought by the competition between skip mode and
CodecNet is evaluated by setting $\balph = 1$ (\textit{CodecNet only}) or
$\balph = 0$ (\textit{Skip only}) configuration. Both configurations are
re-trained starting from the complete system and result in a performance
decrease. In \textit{Skip only} configuration, the system output is directly the
prediction $\prediction$. Since prediction can not explain all the frame to code
the performance saturates at low quality. In \textit{CodecNet only}
configuration, the absence of competition between coding modes results in a rate
increase of 53~\% according to the BD-rate metric. This experiment demonstrates
the benefit of using a competition between skip mode and CodecNet.

\section{Conclusions}

In this paper, a new method for inter frame coding is introduced, based on two
autoencoders: MOFNet and CodecNet. MOFNet role is to compute and convey the
optical flow and a pixel-wise mode selection, allowing to choose between skip
mode and coding through CodecNet. 

The proposed coding scheme performances are illustrated under the CLIC20 P-frame
coding task and it is shown to be competitive HEVC. Moreover, skip mode enables
to learn the optical flow in an actual end-to-end fashion \textit{i.e.} with no
need of a pre-training or a dedicated loss term. 

In future work, we plan to adapt the proposed coding scheme to more complex
video coding tasks such as coding frames with multiple references, both in the
past and in the future. This implies to enhance all sub-networks to leverage as
much information as possible from the references.

\bibliographystyle{IEEEtran}
\bibliography{refs}

\end{document}